\documentstyle[12pt,epsfig]{article}

\newcommand{\hmn}{\mbox{$\vert H_{-\frac{1}{2}-1}\vert^2$}}

\newcommand{\ho}{\mbox{$\vert H_{-\frac{1}{2}0}\vert^2$}}
\newcommand{\hpl}{\mbox{$\vert H_{\frac{1}{2}1}\vert^2$}}
\newcommand{\hoi}{\mbox{$\vert H_{\frac{1}{2}0}\vert^2$}}
\newcommand{\mbmax}{\mbox{{\scriptsize max}}}

\newcommand{\mbos}{\mbox{{\scriptsize off-shell}}}

\def\simgt{\rlap{\lower 3.5 pt \hbox{$\mathchar \sim$}} 
           \raise 1pt \hbox {$>$}}
\def\simlt{\rlap{\lower 3.5 pt \hbox{$\mathchar \sim$}} 
           \raise 1pt \hbox {$<$}}

\catcode`\@=11 
\def\@citex[#1]#2{\if@filesw\immediate\write\@auxout{\string\citation{#2}}\fi
  \def\@citea{}\@cite{\@for\@citeb:=#2\do
    {\@citea\def\@citea{,\penalty\@m}\@ifundefined
       {b@\@citeb}{{\bf ?}\@warning
       {Citation `\@citeb' on page \thepage \space undefined}}%
\hbox{\csname b@\@citeb\endcsname}}}{#1}}
\def\citer{\@ifnextchar [{\@tempswatrue\@citexr}{\@tempswafalse\@citexr[]}}
\relax
\begin{document}
\thispagestyle{empty}

\vspace*{2cm}

\begin{center}
{\bf\Large\sc Infinite Momentum Frame Calculation of}
\vglue .1cm
{\bf\Large\sc Semileptonic Heavy $ \Lambda_{b} \rightarrow \Lambda_{c}$ 
               Transitions}
\vglue .1cm
{\bf\Large\sc including HQET Improvements} 
\vglue 1.2cm
\begin{sc}
Barbara K\"onig$^{a}$,
J\"urgen G.\ K\"orner$^{a,}$\footnote{Supported in part by the BMFT, 
                                     FRG under contract 06MZ865},
Michael Kr\"amer$^{b}$\\ 
\vglue .1cm
and Peter Kroll$^{c}$\\
\vglue 0.5cm
\end{sc}
$^{a}${\it Institut f\"ur Physik, Johannes Gutenberg-Universit\"at, 
           D-55099 Mainz, FRG}

$^{b}${\it Rutherford Appleton Laboratory, Chilton, Didcot, OX11 0QX, 
           UK}

$^{c}${\it Fachbereich Physik, Universit\"at Wuppertal,
           D-42097 Wuppertal, FRG}
\end{center}
\vglue 1.2cm

\begin{abstract}
\noindent
  We calculate the transition form factors that occur in heavy
  $\Lambda$-type baryon semileptonic decays as e.g. in $\Lambda_{b}
  \rightarrow \Lambda_{c}^{+} + l^{-} + \overline{\nu}_{l} $.  We use
  Bauer-Stech-Wirbel type infinite momentum frame wave functions for
  the heavy $\Lambda$-type baryons which we assume to consist of a
  heavy quark and a light spin-isospin zero diquark system. The form
  factors at $ q^{2} = 0 $ are calculated from the overlap integrals
  of the initial and final $\Lambda$-type baryon states.  To leading
  order in the heavy mass scale the structure of the form factors
  agrees with the HQET predictions including the normalization at zero
  recoil.  The leading order $\omega$-dependence of the form factors
  is extracted by scaling arguments. By comparing the model form
  factors with the HQET predictions at ${\cal O}(1/m_{Q})$ we obtain a
  consistent set of model form factors up to ${\cal O}(1/m_{Q})$. With
  our preferred choice of parameter values we find that the
  contribution of the non-leading form factor is practically
  negligible. We use our form factor predictions to compute rates, 
  spectra and various asymmetry parameters for the semi-leptonic decay
  $\Lambda_{b} \rightarrow \Lambda_{c}^{+} + l^{-} + \overline{\nu}_{l} $.
\end{abstract}

\vfill
\newpage

\section{Introduction}
The first evidence of semileptonic $\Lambda_{b}$ had been reported by
the ALEPH and OPAL Collaborations who had seen an excess of correlated
$\Lambda_{s} l^{-} $ pairs over $\Lambda_{s} l^{+} $ pairs (with high
$p_{T}$ leptons) from Z decays \cite{aleph1,opal}.  The $\Lambda_{s}
l^{-} $ excess was readily interpreted as evidence for semileptonic
decays of bottom $\Lambda$-baryons via the chain $\Lambda_{b}
\rightarrow \Lambda_{c} \rightarrow \Lambda_{s} $ \cite{aleph1,opal}.
In the meantime some of the $\Lambda_{c}$ in the event sample have
been fully reconstructed using the decay channel
$\Lambda_{c}^{+}\rightarrow p K^{-} \pi^{+}$ \cite{aleph2}.  Most
recently, the CDF Collaboration \cite{cdf} at the FERMILAB Tevatron
Collider measured the lifetime of the $\Lambda_{b}$ using its
semileptonic decay based on an event sample of 197 $\pm$ 25
reconstructed semileptonic decays. From the experience with s.l.\ 
bottom meson decays, one expects a significant fraction of the s.l.\ 
$\Lambda_{b} \rightarrow \Lambda_{c}^{+} X l^{-} \overline{\nu}_{l}$
transitions to consist of the exclusive mode $\Lambda_{b} \rightarrow
\Lambda_{c}^{+} l^{-}\overline{\nu}_{l}$.  One can be quite hopeful
that fully reconstructed s.l.\ $\Lambda_{b} \rightarrow \Lambda_{c}$
events will become available in the near future.

It is therefore important to study theoretical models for the
$\Lambda_{b} \rightarrow \Lambda_{c}$ transition form factors
including their velocity transfer (or momentum transfer) dependence.
In the heavy meson sector there has been a calculation of the
$B\rightarrow D (D^{*})$ current-induced heavy meson transition form
factors in terms of the Bauer-Stech-Wirbel (BSW) form factor model
which was improved to ${\cal O}(1/m_{Q})$ by comparison with the Heavy
Quark Effective Theory (HQET)\cite{nr}. It is the purpose of this
paper to provide corresponding form factor calculation for the
baryonic $\Lambda_{b} \rightarrow \Lambda_{c}$ transitions using again
BSW type form factors improved by HQET.

Let us briefly review the ${\cal O}(1)$ and ${\cal O}(1/m_{Q})$
structure of the $\Lambda_{b}\rightarrow\Lambda_{c}$ form factors as
predicted by HQET \cite{ggw}.  For that purpose we choose to define
the three vector and axial vector form factors $f_{i}^{V}$ and
$f_{i}^{A}$ by
\begin{eqnarray}\label{fivfia}
 <\Lambda_{c} (v_{2})  \mid V_{\mu} \mid \Lambda_{b} (v_{1}) >
&=& \overline{u} (v_{2}) (
f_{1}^{V} \gamma_{\mu} + f_{2}^{V} v_{1\mu} + f_{3}^{V} v_{2\mu}
                               ) u (v_{1})
\\
 <\Lambda_{c} (v_{2})  \mid A_{\mu} \mid \Lambda_{b} (v_{1}) >
&=& \overline{u} (v_{2}) (
f_{1}^{A} \gamma_{\mu}\gamma_{5} + f_{2}^{A} v_{1\mu}\gamma_{5}
+ f_{3}^{A} v_{2\mu}\gamma_{5}
                               ) u (v_{1})
\nonumber
\end{eqnarray}
In the following we switch to a more generic notation and identify the
labels $b$ and $c$ with 1 and 2, respectively.  In Eq.\ (\ref{fivfia})
we have used velocity covariants to define our form factors as is
appropriate when discussing the ramifications of heavy quark symmetry.
We define the velocity transfer variable $\omega$ by $\omega = v_{1}
\cdot v_{2} $, as usual. We use a conventional state normalization and
normalize our spinors by $\overline{u}u = 2M$.  The ${\cal O}(1)$ HQET
predictions for the form factors read as follows \cite{hkm,hkkt,isgur}
\begin{eqnarray}\label{fivfiaone}
{\cal O}(1)&:& f_{1}^{V} (\omega) = f_{1}^{A} (\omega):= F(\omega)
        \nonumber\\
    & &f_{2}^{V}(\omega) = f_{3}^{V} (\omega) =
       f_{2}^{A}(\omega) = f_{3}^{A} (\omega) = 0
\end{eqnarray}
The ${\cal O}(1)$ reduced form factor $F(\omega)$ satisfies the zero
recoil normalization condition \\$F(\omega = 1) = 1$ \cite{hkkt,isgur}.

Before writing down the ${\cal O}(1/m_{Q})$ corrections we note that
there are two different sources for the ${\cal O}(1/m_{Q})$
corrections that come into play. First one has a local contribution
from the $1/m_{Q}$ corrected HQET current which is proportional to the
binding energy of the $\Lambda_Q$ baryon denoted by $\bar{\Lambda}$ (
$\bar{\Lambda} \approx M_{Q}-m_{Q}\approx600 \,\,\mbox{MeV}$) and to
the ${\cal O}(1)$ reduced form factor $F(\omega)$. Second there is a
nonlocal contribution coming from the kinetic energy term of the
$1/m_{Q}$ corrected HQET Lagrangian. The evaluation of this
contribution brings in a new reduced form factor which will be denoted
by $\eta(\omega)$.\footnote{It is important to realize that there is
  no contribution from the chromomagnetic $1/m_{Q}$ term in the HQET
  Lagrangian in the case of the $\Lambda_b \rightarrow \Lambda_c$
  transitions since the light-side transition is a spin-0 to spin-0
  transition in this case.} Accordingly we have (see e.g.~\cite{KKP})
\begin{eqnarray}
  \label{kkp}
  f_1^V(\omega,m_1,m_2) &=& F(\omega)\; + \;\frac{1}{2}
                            \left[\frac{1}{m_c} + \frac{1}{m_b}\right]\;
                     \left(\eta(\omega) + \bar{\Lambda}F(\omega) \right)
                                                             \nonumber\\
  f_1^A(\omega,m_1,m_2) &=& F(\omega)\; + \;\frac{1}{2}
                             \left[\frac{1}{m_c} + \frac{1}{m_b}\right]\;
                       \left(\eta(\omega) + \bar{\Lambda}F(\omega)
                           \frac{\omega-1}{\omega+1} \right ) \nonumber\\
  f_2^V(\omega,m_1,m_2) &=& \quad f_2^A(\omega,m_1,m_2) =
                     -\frac{1}{m_c}
                      \frac{\bar{\Lambda}F(\omega)}{1+\omega} \nonumber\\
  f_3^V(\omega,m_1,m_2) &=& - f_3^A(\omega,m_1,m_2) =
                     -\frac{1}{m_b}
                      \frac{\bar{\Lambda}F(\omega)}{1+\omega}
\end{eqnarray}
where $\eta(\omega)$ satisfies the zero recoil normalization condition
$\eta(\omega=1)=0$. We have written out the $m_1,m_2$ dependence in
the arguments of the form factors in order to clearly exhibit the
scaling structure of the various contributions in Eq.\ (\ref{kkp}).
Eq.\ (\ref{kkp}) shows that, up to ${\cal O}(1/m_{Q})$, the six form
factors are given in terms of the ${\cal O}(1)$ function $F(\omega)$,
the ${\cal O}(1/m_{Q})$ function $\eta(\omega)$ and the constant
$\bar{\Lambda}$.  One of the entreating features of HQET is that, up
to ${\cal O}(1/m_{Q})$, one retains a zero recoil normalization
condition for the form factors which reads \cite{ggw}
\begin{eqnarray}\label{normv}
f_{1}^{V}(1) + f_{2}^{V}(1) + f_{3}^{V}(1) &=& 1 \nonumber\\
                              f_{1}^{A}(1) &=& 1
\end{eqnarray}
Note that the linear combinations of amplitudes written down in Eq.\ 
(\ref{normv}) are nothing but the vector and axial vector current
s-wave amplitudes, respectively. They give the dominant contributions
at pseudothreshold (or zero recoil) as $\omega\rightarrow 1$.  Put in
a different language, the vector combination and axial vector term in
(\ref{normv}) make up the so-called allowed Fermi and Gamow-Teller
transitions, resp., which are induced by the time component $V_{0}$ of
the vector current and space components $A_{i}$ (i=1,2,3) of the axial
vector current.  It is important to keep in mind that only the s-wave
amplitudes Eq.\ (\ref{normv}) are constrained by HQET to 
${\cal O}(1/m_{Q})$ at zero recoil.

A different but equivalent representation of the ${\cal O}(1) + {\cal
  O}(1/m_{Q})$ HQET result Eqs.\ (\ref{fivfiaone},\ref{kkp}) may be
written down in the form \cite{ggw}\\
${\cal O}(1) + {\cal O}(1/m_{Q}):$
\begin{eqnarray}\label{georgi}
f_{1}^{V} (\omega,m_1,m_2) &=&\quad f_{1}^{A} (\omega,m_1,m_2)
\left(1+\left[\frac{1}{m_1}+\frac{1}{m_{2}}\right]
\frac{\bar{\Lambda}}{\omega + 1}\right)
\nonumber\\
f_{2}^{V}  (\omega,m_1,m_2) &=& \quad f_{2}^{A} (\omega,m_1,m_2)
= - f_{1}^{A}  (\omega,m_1,m_2)
\frac{1}{m_{2}} \frac{\bar{\Lambda}}{\omega + 1}
\nonumber\\
f_{3}^{V} (\omega,m_1,m_2) &=& - f_{3}^{A} (\omega,m_1,m_2)
= - f_{1}^{A} (\omega,m_1,m_2)
\frac{1}{m_{1}} \frac{\bar{\Lambda}}{\omega + 1}
\end{eqnarray}
where the ${\cal O}(1/m_Q)$ zero recoil normalization conditions now
reads
\begin{equation}\label{0recoilnorm}
f_{1}^{A}(\omega, m_1, m_2) = 1
\end{equation}
It is not difficult to see that the two representations (\ref{kkp})
and (\ref{georgi}) are equivalent at ${\cal O}(1/m_{Q})$ but start to
be different at ${\cal O}(1/m_{Q}^2)$. The representation
(\ref{georgi}) is somewhat simpler in that there is only the one
$\omega$-dependent function $f_{1}^{A}(\omega, m_1, m_2)$. The
representation (\ref{kkp}) has the advantage that the various ${\cal
  O}(1)$ and ${\cal O}(1/m_{Q})$ contributions remain clearly
identified. It is for this reason that we shall work with the
representation (\ref{kkp}) in the following.

Let us now turn to our model calculation to determine the
$\omega$-dependence of the ${\cal O}(1)$ reduced form factor
$F(\omega)$ and the $[{\cal O}(1) + {\cal O}(1/m_{Q})]$ form
factor $f_{1}^{A} (\omega)$, or, in the representation (\ref{kkp}), of
the ${\cal O}(1/m_Q)$ reduced form factor $\eta(\omega)$.  We employ
the method introduced by Neubert and Rieckert \cite{nr} which allows
one to determine the hadronic form factors as functions of the scaling
variable $\omega$ once they are known at $q^2 = 0$. The idea of
Neubert and Rieckert is as follows: For general momentum transfer
squared $q^2$ the relation between $q^2 $ and the scaling variable
$\omega$ is given by
\begin{equation}
\omega = v_{1}\cdot v_{2}= \frac{M_{1}^2 + M_{2}^2 - q^2 }{2 M_{1} M_{2}}
\end{equation}
Neubert and Rieckert propose to compute quark model form factors at
$q^2 =0$ where one has \footnote{ The maximum recoil point $q^2 =0$ is
  privileged in the IMF quark model approach of BSW \cite{bsw}.  At
  $q^2 =0$ the IMF overlap integrals allow for a specific
  interpretation in terms of space integrals of the 'good' current
  components that are the charges of a broken collinear symmetry at
  infinite momentum. This means that, in the limit of a strict
  collinear symmetry combining spin and flavour (i.e. an $SU(4)_{W}$
  symmetry acting on two spin states and two appropriate quark
  flavours), the normalized overlap integrals of the 'good' current
  components are generators of this collinear spin-flavour group. A
  drawback of the IMF at $q^2 =0$ is that it cannot be related to a
  frame of finite momentum by any Lorentz transformation.}
\begin{equation}
\omega(q^2=0)=
\frac{1}{2}\left(\frac{M_{1}}{M_{2}}+\frac{M_{2}}{M_{1}}\right)
\end{equation}
Then by varying the ratio $M_1/M_2$ at the point $q^2 =0$ in the quark
model calculation one can extract the form factor values for all
values of the scaling variable $\omega \geq 1$ provided one is using
appropriate scaling form factors.  The appropriate scaling form
factors of the quark model are identified by comparison with the HQET
structure Eq.\ (\ref{fivfiaone}) or (\ref{kkp}). Explicitly, we shall
relate the form factors at $q^2 = 0$ to overlap integrals as is done
in the BSW scheme. Using a diquark model for the infinite momentum
frame (IMF) wave function \cite{bsw}, the overlap integrals can be
evaluated. It follows trivially that the ${\cal O}(1)$ form factors
$f_{1}^{V} = f_{1}^{A} = F(\omega)$ are normalized for $M_1 = M_2$ or
$\omega = 1$ since they correspond to an overlap integral which is
normalized to one, i.e. $< M_{Q} | M_{Q} >=1$ for identical hadron
states.  The overlap integrals and thus the quark model form factors
can be expanded w.r.t.\ the inverse heavy quark masses.  One can
identify the zeroth and first order terms in this expansion with the
same expansion in HQET and thereby compute the $\omega$-dependence of
the ${\cal O}(1)$ and ${\cal O}(1/m_Q)$ reduced form factors that appear
in Eqs.\ (\ref{kkp}) or (\ref{georgi}) by varying the mass ratio
$M_1/M_2$.

\section{Infinite Momentum Frame Wave Functions}
As explained before we shall employ the approach of BSW \cite{bsw} to
calculate form factors at $q^2 = 0 $ in terms of relativistic bound
state wave functions in the infinite momentum frame.  In the
relativistic BSW approach the hadrons are described as relativistic
bound states of a heavy active quark $Q_1$ and a heavy or light
spectator state, which, in our case is a spin-isospin zero light
diquark state.  A relativistic bound state of a quark-diquark pair in
the IMF is written as
\begin{eqnarray}\label{relbs}
|{\bf P}, M_i; J,J_{z}> &=& \sqrt{2} (2 \pi)^{3/2} \sum_{s_1 s_2}
\int {\rm d}^3 p_1 {\rm d}^3 p_2 \delta^{3} ({\bf P} - {\bf p}_{1} -
 {\bf p}_{2})
\\ &&
\Phi_i^{J , J_z} ({\bf p}_{1 \perp}, x_1; s_1, s_2)
\,a_1^{\dagger}({\bf p}_1, s_1) a_2^{\dagger}({\bf p}_2, s_2)\, | 0 >
\nonumber
\end{eqnarray}
where $a_1^{\dagger}({\bf p}_1 , s_1)$ denotes the creation operator
for the heavy quark and $ a_2^{\dagger}({\bf p}_2 , s_2)$ the creation
operator for the light diquark and where ${\bf p}_{1} ({\bf p}_{2})$,
$s_{1} (s_{2})$ represent the momentum and spin of the heavy quark
(light diquark), respectively.  The fraction of the longitudinal
momentum carried by the active heavy quark $Q_1$, is denoted by $x_1 =
p_{1 z}/P$, ${\bf p}_{1 \perp} = ( p_{1x},p_{1y}) $ is the relative
transverse momentum of the active heavy quark. We use a conventional
state nor\-ma\-li\-za\-tion \mbox{$ < {\bf P}' | {\bf P} > = 2 P^0 (2\pi)^3
  \delta^3 ({\bf P} - {\bf P}')$} so that 
\begin{equation}
\sum_{s_1 s_2} \int {\rm d}^2 p_{\perp} {\rm d}x \, 
        |\Phi_{i}^{J J_z}({\bf p}_{\perp}, x; s_1, s_2)|^2\;= \;1.  
\end{equation}
In the following we suppress spin labels.  The
IMF heavy ground state baryon wave function is constructed in complete
analogy to the heavy ground state meson case. The light antiquark in
the meson case is substituted by a light diquark in the baryon case.
In our definition the wave function describes a quark-diquark Fock
state for the $\Lambda$-type baryon in which the spin degrees of
freedom decouple from the momentum.  The two light degrees of freedom
are treated as a single quasi-elementary constituent and are
represented by a spin-isospin zero diquark with $[ud]$ quantum
numbers.  Because the diquark always shows up as a spectator in the
overlap integrals it is of no significance whether or not it is a true
bound state. Colour indices are omitted for convenience.  $\Phi_i (x_1
, p_{\perp})$ denotes the hadronic IMF (null plane) wave function
normalized to one where $i = b, c$. This momentum space wave function
is assumed to factorize into its longitudinal and transverse momentum
dependence, i.e.\ into its $x$- and $p_{\perp}$-dependence as usual.
Large transverse momenta are assumed to be strongly suppressed by
introducing an exponential cut-off. Such a picture is corroborated by
many observations in inclusive processes. One thus has
\begin{equation}\label{separation}
\Phi_i (x_1 , p_{\perp}) = N_i \phi_{i}(x_1) \exp \{ - b^2_i p_{\perp}^2 \}
\end{equation}
$N_i$ is a normalization factor whose value is fixed once the
$x_1$-dependence of $\phi_i(x_1)$ is specified.  The oscillator
parameter $b_i$ characterizes a soft process scale below which there
is no suppression by the wave function.  It may be subject to mass
corrections. Therefore we make the ansatz $b_i= b + \bar{b}/M_i$.
Interpreting the transverse momenta as Fermi motion of the baryon
constituents, the oscillator parameter $b_i$ is adjusted such that
realistic mean $p_{\perp}$'s are obtained, i.e.\ a mean $p_{\perp}$ is
of the order of a few hundred MeV. From this consideration we expect
a value of order $1-2$~GeV$^{-1}$ for $b$. Guided by results for
mesonic decays \cite{fau} we expect $\bar b$ to be about 0.1.

The $x_1$-dependence of the hadronic wave function $\phi_i(x_1)$ is
controlled by the long distance behaviour of QCD. The calculation of
the $x_1$-dependence would require nonperturbative methods as e.g.\ 
lattice gauge theories. As there are no nonperturbative results yet we
have to rely on educated guesses. Most appropriate for our purposes is
the wave function
\begin{equation}\label{nm}
\phi_i (x_1) = \overline{N}_i x_1 ^n (1-x_1)^m
\exp\{- b^2_i M_i^2 (x_1 - x_{i0})^2 \}
\end{equation}
This wave function is a generalization of the meson wave function
proposed by Bauer, Stech and Wirbel \cite{bsw}.  It has already been
used for the description of heavy baryons in the large recoil region
\cite{kokro} as well as for light baryons in a quark-diquark model
\cite{krollschwei}.

The wave function (\ref{nm}) exhibits a pronounced maximum at $x_{i0}
= 1 - \alpha_i/M_i$ where $\alpha_i$ is the difference between the
masses of the heavy hadron and the heavy quark, i.e. $\alpha_i = M_i -
m_i$ which can be expanded over inverse powers of $M_i$ \cite{fal}. We
take into account only the the first two terms of this series
$\alpha_i= \alpha + \bar{\alpha}/M_i$.  In the zero binding
approximation $\alpha$ is approximately equal to the diquark mass and
lies in the range of $0.5 - 1.0$ GeV. The size of the correction term
$\bar{\alpha}$ has been estimated from baryonic QCD sum rules to
amount to $0.1-0.3$~GeV$^2$~\cite{col}.  Finally, $\overline{N}_i$ is a 
further normalization constant fixed by the requirement
\begin{equation}\label{normNi}
\int_{0}^{1} {\rm d}x_1 \phi_{i}^{*} (x_1) \phi_i (x_1) = 1
\end{equation}
which depends on the values chosen for the endpoint powers $n$ and
$m$.

The behaviour of the wave function (\ref{nm}) in the endpoint regions
$x_1 \rightarrow 0$ and $x_1 \rightarrow 1$ is controlled by the power
dependent terms $x_{1}^{n}$ and $(1- x_{1})^{m}$. As the endpoint
behaviour of the light diquark system can be expected to be the same
as that of the heavy quark the choice $n = m$ seems to be a natural
one. We note that for light baryons this choice is not optimal
\cite{krollschwei}.  Finally, in our numerical work we take $n = m =
1/2$. This is suggested by the similarity of the baryonic heavy
quark-light diquark system and the mesonic heavy quark-light antiquark
system for which this choice has been found to be appropriate
\cite{bsw}.  For the sake of comparison we also compute numerical
results for the powers $n = m = 1$. We mention that the polarization
predictions of our model are only weakly dependent on the choice of
$n$ and $m$, whereas the rate prediction does depend on the choice of
$n$ and $m$.

\section{Model form factors}
To leading order in the IMF momentum P one finds two relations for the
heavy baryon decay form factors.  These two relations correspond to
the 0- and 3-component of the transition current which represent
leading order contributions in the IMF momentum P and are thus termed
'good'. The 1- and 2- components are 'bad', because in the latter,
particles moving with $ x < 0$ or $ x > 1$ cannot be excluded. For
these the extra powers of P in the denominator can be compensated by
similar factors in the numerator from the matrix elements of $J_{1}$
and $J_{2}$, thus mimicking a constant behaviour though they may hide
terms proportional to P as P$\rightarrow \infty$.  A more
phenomenological argument why we have decided to keep only first order
expressions in the matrix elements is that many approximations were
made in our parton model approach.  First of all, though we set $x =
p_{1 z}/P$, there may be components of the heavy quark momentum
perpendicular to the z-direction of ${\bf P}$.  Such transverse
momenta, as well as off-shell effects, the light cone factorization in
$x$ and $p_{\perp}$, which is not rotationally invariant, and the
decoupling of spin and orbital momenta lead to modifications of order
$1/P$. Despite of this the bad current relations are in agreement with
HQET as we show in an Appendix, there is only a little difficulty with
the reduced form factor $\eta(\omega)$.  Our subsequent analysis is
based on the good current components only.

The good relations between the form factors corresponding to
the 0- and 3-components of the current transitions at
$q^2 = 0$ read
\begin{eqnarray}
\label{0and3a}
&&f_{1}^{V}(\omega, M_1, M_2) \,+\, \frac{1}{2}\left (\frac{1}{M_1}
+ \frac{1}{M_2}\right ) \, \left(
M_2 f_{2}^{V}(\omega, M_1, M_2) \, +\,
M_1 f_{3}^{V}(\omega, M_1, M_2) \right) \nonumber\\
&&= I(\omega, M_1, M_2)
\end{eqnarray}
\begin{eqnarray}
\label{0and3b}
&&f_{1}^{A}(\omega, M_1, M_2) \,+\, \frac{1}{2}\left (\frac{1}{M_1}
- \frac{1}{M_2}\right ) \, \left(
M_2 f_{2}^{A}(\omega, M_1, M_2) \, +\,
M_1 f_{3}^{A}(\omega, M_1, M_2) \right) \nonumber\\
&&= I(\omega, M_1, M_2)
\end{eqnarray}
where $I(\omega, M_1, M_2) $ is an overlap integral between the initial
and final state baryons and is given by
\begin{equation}\label{I(y)}
I(\omega, M_1, M_2) =
               \int_{0}^{1}{\rm d}x_1 \phi_2^{*} (x_1) \phi_1 (x_1)
\end{equation}
Note that in the elastic case $M_1 = M_2$ (which implies
$\omega = 1$ at $q^2 =0$) one reads off the normalization
conditions $f_{1}^{V}(1) + f_{2}^{V}(1) +f_{3}^{V}(1) = 1$ and
$f_{1}^{A} (1) = 1$ from (\ref{separation})--(\ref{0and3b}).
We emphasize, though, that Eqs.\ (\ref{separation})--(\ref{0and3b})
imply no normalization condition for $M_1 \not= M_2$ at $\omega = 1$.

In order to extract further information from
Eqs.\ (\ref{separation})--(\ref{0and3b}) we have to expand the form
factors $f_{i}^{V}(\omega, M_1, M_2)$ and $f_{i}^{A}(\omega, M_1, M_2)$,
and the overlap function $I(\omega, M_1, M_2) $ into
appropriate scaling functions that depend on the
scaling variable $\omega$ only.

Let us first expand the overlap function $I(\omega,M_1 ,M_2 )$
into inverse powers of the heavy baryon mass 1/$M_i$, i.e.
\begin{eqnarray}\label{Itilde}
I(\omega,M_1 ,M_2 ) &=& I^{(0)}(\omega) +
\left[\frac{1}{M_1} + \frac{1}{M_2}\right] I^{(1)}(\omega) + ...
\nonumber\\
&:=& I^{(0)}(\omega)
\left(
1+\left[
\frac{1}{M_1}+\frac{1}{M_2}
  \right]\tilde{I}^{(1)}(\omega) + ...
\right)
\end{eqnarray}
where
\begin{eqnarray}\label{itildedef}
I^{(0)}(\omega) &=&
\sqrt{\frac{1}{\omega}}^{2m+1}
\exp{\{ - \kappa^2\frac{\omega -1}{2\omega} \}}
\cdot
\frac{H_{2m}(\kappa\sqrt{\frac{\omega +1}{2\omega}})}{H_{2m}(\kappa)}\\
\tilde{I}^{(1)}(\omega) &=& \;\frac{\omega -1}{\omega}
                            \left[ \frac{2m+1}{2} \frac{\bar{b}}{b}
                            -\sqrt{2}\kappa (\bar{\alpha} b + \alpha \bar{b})
                            + \kappa \alpha \bar{b}/(\sqrt{2}\omega)\right]
                                                                  \nonumber\\
                        & & +\frac{n}{\sqrt{2}b}
                           \left[ \frac{H_{2m+1}(\kappa)}{H_{2m}(\kappa)} -
        \sqrt{\frac{2}{\omega(\omega+1)}} \frac{
           H_{2m+1}(\kappa\sqrt{\frac{\omega+1}{2\omega}})}
           {H_{2m}(\kappa\sqrt{\frac{\omega+1}{2\omega}})} \right]
                                                                   \nonumber\\
    & &  - \frac{1}{\sqrt{2}} (\bar{\alpha} b + \alpha \bar{b}) \left[
          \frac{H'_{2m}(\kappa)}{H_{2m}(\kappa)} -
        \sqrt{\frac{2}{\omega(\omega+1)}}\frac{
         H'_{2m}(\kappa\sqrt{\frac{\omega+1}{2\omega}})}
        {H_{2m}(\kappa\sqrt{\frac{\omega+1}{2\omega}})} \right]
                                                                   \nonumber\\
    & &  + \frac{1}{2} \alpha \bar{b} \frac{\omega-1}{\sqrt{\omega^3(\omega+1)}}
        \frac{H'_{2m}(\kappa\sqrt{\frac{\omega+1}{2\omega}})}
        {H_{2m}(\kappa\sqrt{\frac{\omega+1}{2\omega}})}.
\end{eqnarray}
Furthermore we have introduced the following abbreviations
\begin{eqnarray}\label{h2m}
H_{l}(x)& = &
\int_{- x }^{\infty} {\rm d}z \left(z + x \right)^{l} e^{- z^2} \nonumber\\
H'_l(x) &=& {\rm d} H_l(x)/{\rm d}x \nonumber\\
\kappa &=& \sqrt{2} \alpha b
\end{eqnarray}
Note that the scaling functions $I^{(0)}(\omega)$ and
$\tilde{I}^{(1)}(\omega)$ obey the normalization conditions
$I^{(0)}(1) = 1$ and $\tilde{I}^{(1)}(1) = 0$.

It is clear that the two constraint relations
(\ref{0and3a},\ref{0and3b}) do not suffice to determine the six
unknown form factors. However, when one inserts the HQET relations
(\ref{kkp}) or (\ref{georgi}) into the determining relations one can
then solve for the reduced HQET form factors. Using e.g. the
representation (\ref{kkp}) for the HQET expansion of the form factors
one can then solve the two $q^2=0$ relations Eqs.\ 
(\ref{0and3a},\ref{0and3b}) to obtain
\begin{eqnarray}\label{solution}
{\cal O}(1):  &&   F(\omega) = I^{(0)}(\omega)
                                                              \nonumber\\
{\cal O}(1/m_Q):&& \eta(\omega)\; = I^{(0)}(\omega)
                      \left[ 2 \tilde{I}^{(1)}(\omega)
                    - \bar {\Lambda}\frac{\omega - 1}{\omega + 1} \right]
\end{eqnarray}
It is quite evident that up to order ${\cal O}(1/m_Q)$ the solution
(\ref{solution}) satisfies the zero recoil normalization condition
(\ref{normv}) for the form factors since $F(1) = 1$ and $\eta(1) = 0$.
Note that $F$ depends on $\kappa$ only, i.e.\ on the product of the
two model parameters $\alpha$ and $b$ whereas $\eta$ depends on all
the parameters, $\alpha, \bar{\alpha}, b, \bar{b}$.

It is noteworthy that the $q^2=0$
constraint relations (\ref{0and3a},\ref{0and3b}) provide no
 constraint on the HQET parameter
$\bar{\Lambda}$.
For consistency reasons we fix
\begin{equation}\label{lambdabar}
\bar{\Lambda} = \alpha
\end{equation}

Of interest is also the slope of the ${\cal O}(1)$ reduced form 
factor $F(\omega)$ at $\omega = 1$
which can in fact be obtained in closed form. In terms of the usual
$\omega \simeq 1$ parametrization
\begin{equation}\label{param}
F(\omega) = 1 - \rho^2 (\omega - 1)+ ...
\end{equation}
one finds
\begin{equation}\label{slope}
\rho^2 = \frac{2m + 1}{2} + \frac{1}{2}\kappa^2 + \frac{1}{4}\kappa
\frac{H'_{2m}(\kappa)}{H_{2m}(\kappa)}
\end{equation}
\section{Numerical Results}
We are now in a position to discuss the numerical
implications of the IMF quark model in terms of decay
spectra, decay rates and asymmetry parameters.

In Fig.\ 1 we plot our model predictions for the $\omega$-dependence
of the ${\cal O}(1)$ reduced form factor $F(\omega)$ for the two
values $n=m=1/2$ and $n=m=1$. 
\begin{figure}[t]

\hspace*{12mm}
\epsfig{file=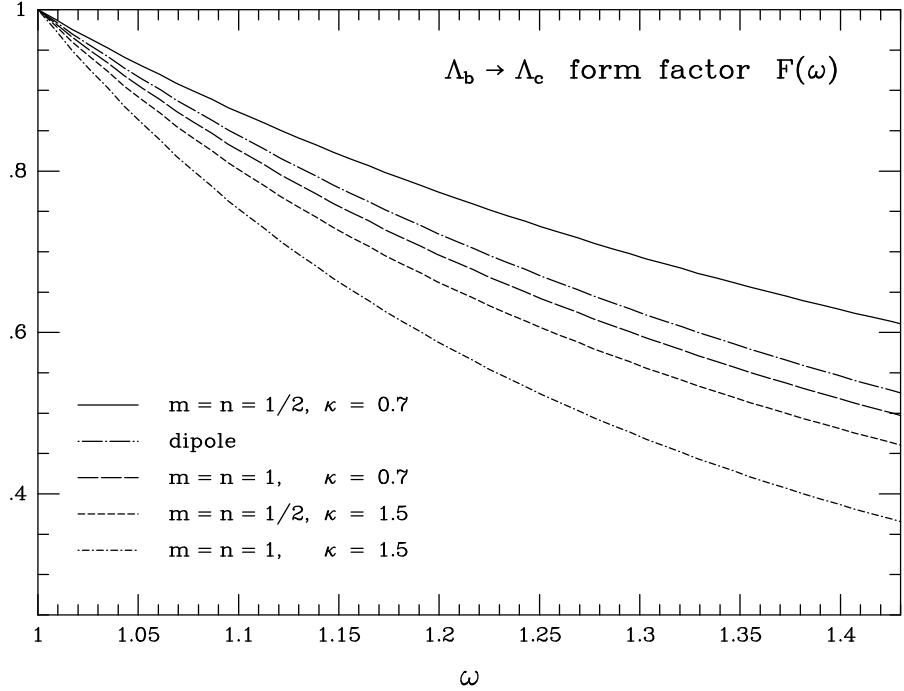,bbllx=10pt,bblly=60pt,bburx=550pt,bbury=770pt,%
        width=9.5cm,height=13.cm,angle=-90}

\caption[dummy]{\label{fig_1} \it Dipole form factor and IMF
  quark model predictions for the ${\cal O}(1)$ reduced form factor
  $F(\omega)$ as functions of $\omega = v_1 \cdot v_2$ $( n=m=1/2, 1;\;\;
  \kappa= 0.7, 1.5)$.}

\end{figure}
In both cases $F(\omega)$ is evaluated for either value of $\kappa$,
0.7 and 1.5. For $\alpha \simeq 0.5 - 0.6$~GeV the latter value of
$\kappa$ corresponds to $b=1.7 - 2.1$ GeV$^{-1}$ which is
characteristic of light baryons \cite{krollschwei,krollguo}.
$\kappa=0.7$, on the other hand, corresponds to $b=0.8 - 1.0$
GeV$^{-1}$ which implies a $\Lambda_b$ radius about half as large as
that on of light baryons. A value of about 1 GeV$^{-1}$ for the
parameter $b$ seems realistic to us.  For the sake of comparison we 
also plot the $\omega$-dependence of a dipole-behaved form factor
$F^{\,\mbox{{\scriptsize dipole}}}(\omega)$ which is appropriately
normalized to one at zero recoil. The normalized dipole form factor is
given by \footnote{ By rewriting (\ref{dipole}) in terms of the
  momentum transfer variable $q^2$ one recovers the familiar dipole
  representation $F^{\,\mbox{{\scriptsize dipole}}}(q^2)= N(q^2)(1 -
  q^2/m_{FF}^{2})^{-2}$ where $N(q^2)$ normalizes the dipole form
  factor to one at the zero recoil point $q^2 = (M_1 - M_2)^2$.}
\begin{equation}\label{dipole}
F^{\,dipole}(\omega)= \left (1 + \frac{2M_1M_2(\omega - 1)}{m_{FF}^{2} -
(M_1 - M_2)^2}\right )^{-2}
\end{equation}
As the form factor mass $m_{FF}$ in the dipole form factor we take the
expected mass value of the $J^{P} = 1^-$ ($b\bar c$) vector meson,
i.e. $m_{FF}=6.34\,\, \mbox{GeV}$.  For the $\Lambda_b$ and
$\Lambda_c$ masses we take $M_1 = M_{\Lambda_b} = 5.621
\,\,\mbox{GeV}$ \cite{cdf} and $ M_2 = M_{\Lambda_c} = 2.285
\,\,\mbox{GeV}$.

The IMF quark model form factors $F(\omega)$ fall more quickly than
the dipole form factor except for our preferred choice $n=m=1/2$ and 
$\kappa=0.7$. We mention that the QCD sum rule analysis of
\cite{Grozin} results in a form factor behaviour which is well
approximated by 
\begin{equation}
F(\omega)=\frac{2}{\omega +1}
             \exp{\left [-(2\rho^2-1)\frac{\omega-1}{\omega+1}\right ]}
\end{equation}
with $\rho \simeq 1$. Thus, the form factor of \cite{Grozin} is even
slightly flatter than our preferred form factor. The fall-off
behaviour of the various form factors $F(\omega)$ in Fig.\ 1 can be
conveniently characterized by comparing the charge radii $\rho^2$ of the form
factors defined in Eq.\ (\ref{param}).  We obtain (see Eq.\ 
(\ref{slope}))
\begin{equation}
\rho^2 = \left\{ \!
\begin{array}{l}
               1.44 \quad \mbox{IMF model ($n=m=1/2$, $\kappa=0.7$)}
            \\ 3.04 \quad \mbox{IMF model ($n=m=1$, $\kappa=1.5$)}
            \\ 1.77 \quad \mbox{dipole model}
\end{array} \right.
\end{equation}

To set the scale of the slope we remind the reader that the normalized
dipole form factor in the infinite mass limit is given by $( 1 +
\frac{1}{2} (\omega - 1) )^{-2}$ and thus has a slope $\rho^2 = 1$. We
mention that in the heavy meson case slope values between $\rho^2 = 1$
and $\rho^2 = 2$ are being discussed in the literature. As mentioned
before the slope of the sum rule form factor of \cite{Grozin} is also
$\rho^2\simeq 1$.

A first measurement of the slope parameter in the transition
$\Lambda_b \rightarrow \Lambda_c$ has been reported in \cite{Bertini}.
The value quoted is
\begin{equation}
\rho^2=1.81^{+0.70}_{-0.67}({\rm stat.})\pm 0.32({\rm syst.}).
\end{equation}
The IMF model value $(n=m=1/2,\ \kappa=0.7)$ and the dipole model value 
both lie wirhin the error bars of this measurement. 

\begin{figure}[t]

\hspace*{12mm}
\epsfig{file=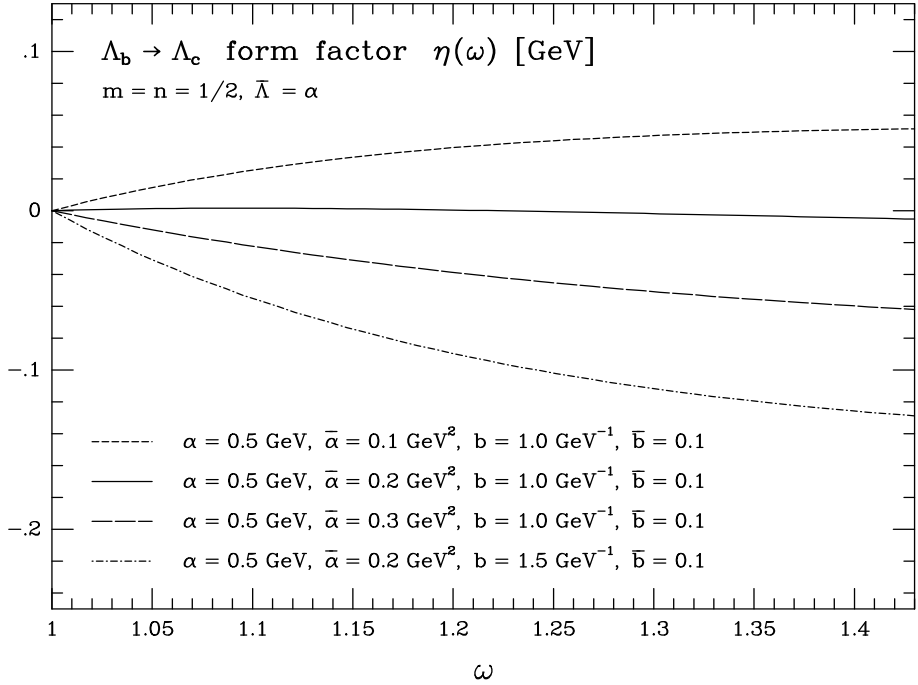,bbllx=10pt,bblly=60pt,bburx=550pt,bbury=770pt,%
        width=9.5cm,height=13cm,angle=-90}

\caption[dummy]{\label{fig_2} \it IMF quark model predictions for the 
  ${\cal O}(1/m_Q)$ reduced form factor $\eta(\omega)$ as functions of
  $\omega = v_1 \cdot v_2$ $\;(m=n= 1/2)$.}

\end{figure}
\begin{figure}[h,t]

\hspace*{12mm}
\epsfig{file=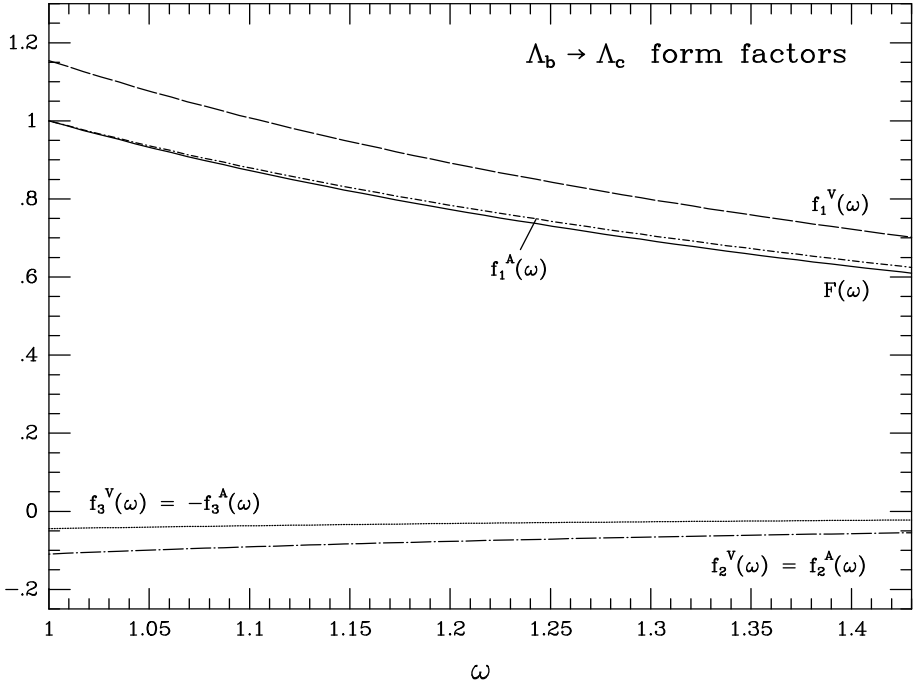,bbllx=10pt,bblly=60pt,bburx=550pt,bbury=770pt,%
        width=9.5cm,height=13cm,angle=-90}

\caption[dummy]{\label{fig_3} \it ${\cal O}(1) + {\cal O}(1/m_Q)$ IMF 
  quark model form factors as functions of $\omega = v_1 \cdot v_2$
  $\;(n=m=1/2)$. ${\cal O}(1)$ form factor $F(\omega)$ is also shown.}

\end{figure}

As explained in Sec.\ 2 our preferred choice for the endpoint
behaviour of the heavy quark-light diquark IMF wave function is $n = m
= 1/2$. In the following we shall no longer discuss the choice $n = m
= 1$, in particular as the slope of the corresponding ${\cal O}(1)$
form factor at $\omega =1$ seems unrealistically big for any reasonable
value of $\kappa$.

In Fig.\ 2 we show our results for the ${\cal O}(1/m_Q)$ reduced form
factor $\eta(\omega)$ obtained with various sets of the model
parameters. It turns out that $\eta(\omega)$ is very small in the
$\Lambda_b$ decay region and, depending on the parameter values, it
can be positive or negative. Judging from the smallness of the
non-leading reduced form factor $\eta(\omega)$ it will not be an easy task to
measure it. Our results for the form factor $\eta(\omega)$ for for 
$b=1.0$ GeV$^{-1}$ and $\bar{\alpha}=0.2\,$ GeV$^2$ are close to those
obtained from QCD sum rules \cite{dai96}.

Inserting the reduced form factors $F(\omega)$ and $\eta(\omega)$ into
(\ref{kkp}), we evaluate the form factors $f_i^{V,A}$ to ${\cal
  O}(1/m_Q)$. We use particle masses throughout in Eq.\ 
(\ref{kkp}). This is perfectly legitimate since the difference between
particle and quark masses is an ${\cal O}(1/m_{Q}^{2})$ effect.  In
Fig.\ 3 we exhibit the $\omega$-dependence of the form factors
$f_i^{V,A}$ for our preferred set of parameters $m=n=1/2$,
$\alpha=0.5$~GeV and $b=1.0$~GeV$^{-1}$ as well as, according to the discussion
in Sect.\ 2, $\bar{\alpha}=0.2$~GeV$^2$ and $\bar{b}=0.1$. Looking at 
Fig.\ 2 this choice of parameters gives a non-leading reduced form factor $\eta(\omega)$ 
which is practically zero. The $1/m_Q$ corrections to the form factors can be
seen to be quite moderate as the comparison with the ${\cal O}(1)$
reduced form factor $F(\omega)$ shows (remember that $f_{2}^{V} =
f_{3}^{V} = f_{2}^{A} = f_{3}^{A} = 0$ at ${\cal O}(1)$). The axial
form factor $f_1^{A}(\omega)$ is predicted to be rather similar to its
${\cal O}(1)$ counterpart $F(\omega)$: the difference amounts to
maximally $\approx 3\%$ at $\omega_{\mbmax}$.  The form factors
$f_{2}^{V}$, $f_{3}^{V}$, $f_{2}^{A}$ and $f_{3}^{A}$ acquire slight
non zero values at ${\cal O}(1/m_Q)$ which are largest at zero recoil.
They never amount to more than $\approx 10\%$ of $f_{1}^{A}$ though.

To proceed further let us first state the linear relations between the
"velocity" form factors defined in Sec.\ 1 and the helicity amplitudes
that enter into the formulae for physical observables.
One has \cite{bkkz}
\begin{eqnarray}\label{ha}
\sqrt{q^2}H^{V,A}_{\frac{1}{2}0}&=&
\sqrt{2M_1M_2(\omega\mp 1)}
\Big(
(M_1\pm M_2)f^{V,A}_1\pm M_2(\omega\pm 1)f^{V,A}_2\nonumber\\
& & \pm M_1(\omega\pm 1)f^{V,A}_3 \Big)\nonumber\\
H^{V,A}_{\frac{1}{2}1}&=&
-2\sqrt{M_1M_2(\omega\mp 1)}f^{V,A}_1
\end{eqnarray}
where $H_{\lambda_2,\lambda_W}^{V,A}$ are the helicity amplitudes for
the vector ($V$) and axial vector ($A$) current induced $1/2^+
\rightarrow 1/2^+ + W^{-}_{\mbos}$ transitions ($\lambda_2$ and
$\lambda_W$ are the helicities of the final state baryon and
$W^{-}_{\mbos}$, resp.). The upper and lower signs in (\ref{ha}) stand
for the vector ($V$) current and axial vector ($A$) current
contributions, resp., where the total helicity amplitude is given by
\begin{equation}
H_{\lambda_2\lambda_W} = H_{\lambda_2\lambda_W}^{V}
 - H_{\lambda_2\lambda_W}^{A}
\end{equation}
for a left-chiral $\gamma_\mu(1-\gamma_5)$ transition. The remaining
helicity amplitudes are related to the above two helicity amplitudes
                                                 by parity. One has
\begin{equation}
H_{-\lambda_2-\lambda_W}^{V,A} = \pm H_{\lambda_2\lambda_W}^{V,A}
\end{equation}
For the differential decay rate one then obtains\cite{kk}
\begin{eqnarray}\label{ddr}
\frac{\mbox{d}\Gamma}{\mbox{d}\omega} & = & \frac{G^2}{(2\pi)^3}
\vert V_{bc}\vert^2\frac{q^2 p M_2 }{12 M_1}\left (
\vert H_{\frac{1}{2}1}\vert^2 +
\vert H_{-\frac{1}{2}-1}\vert^2 +
\vert H_{\frac{1}{2}0}\vert^2 +
\vert H_{-\frac{1}{2}0}\vert^2 \right ) \nonumber \\
& := & \frac{\mbox{d}\Gamma_{T_+}}{\mbox{d}\omega} +
       \frac{\mbox{d}\Gamma_{T_-}}{\mbox{d}\omega} +
       \frac{\mbox{d}\Gamma_{L_+}}{\mbox{d}\omega} +
       \frac{\mbox{d}\Gamma_{L_-}}{\mbox{d}\omega}
\end{eqnarray}
where $p$ is the CM momentum of the $\Lambda_c$ ($ p =
M_2\sqrt{(\omega + 1)(\omega - 1)}$ ).  In the second line of Eq.\ 
(\ref{ddr}) we have defined rates into particular helicity components
through $\Gamma_{T_+} ( \propto \vert H_{\frac{1}{2}1} \vert ^{2})$,
$\Gamma_{T_-} ( \propto \vert H_{-\frac{1}{2}-1}\vert ^{2})$,
$\Gamma_{L_+} ( \propto \vert H_{\frac{1}{2}0}\vert ^{2})$ and
$\Gamma_{L_-} ( \propto \vert H_{-\frac{1}{2}0}\vert ^{2})$ where $T$
and $L$ denote the transverse and longitudinal components of the
current transition.

In Fig.\ 4 we plot the velocity transfer dependence of the
differential decay rate for the IMF quark model with and without
$1/m_Q$ corrections and again compare them to the predictions of the
dipole model. To be definite we have chosen $V_{bc} = 0.044$. For
other values of $V_{bc}$ the decay rate is to be scaled by
$(V_{bc}/0.044)^2$. As would
be anticipated from the comparison of the form factors in Fig.\ 1 the
differential rates of the IMF quark model are larger than the dipole
model rates. The IMF quark model spectrum peaks at larger values of
$\omega$ than the dipole model spectrum. In the case of the IMF quark
model, the $1/m_Q$ corrections tend to slightly increase the 
${\cal O}(1)$ rates.
\begin{figure}[htbp]

\hspace*{12mm}
\epsfig{file=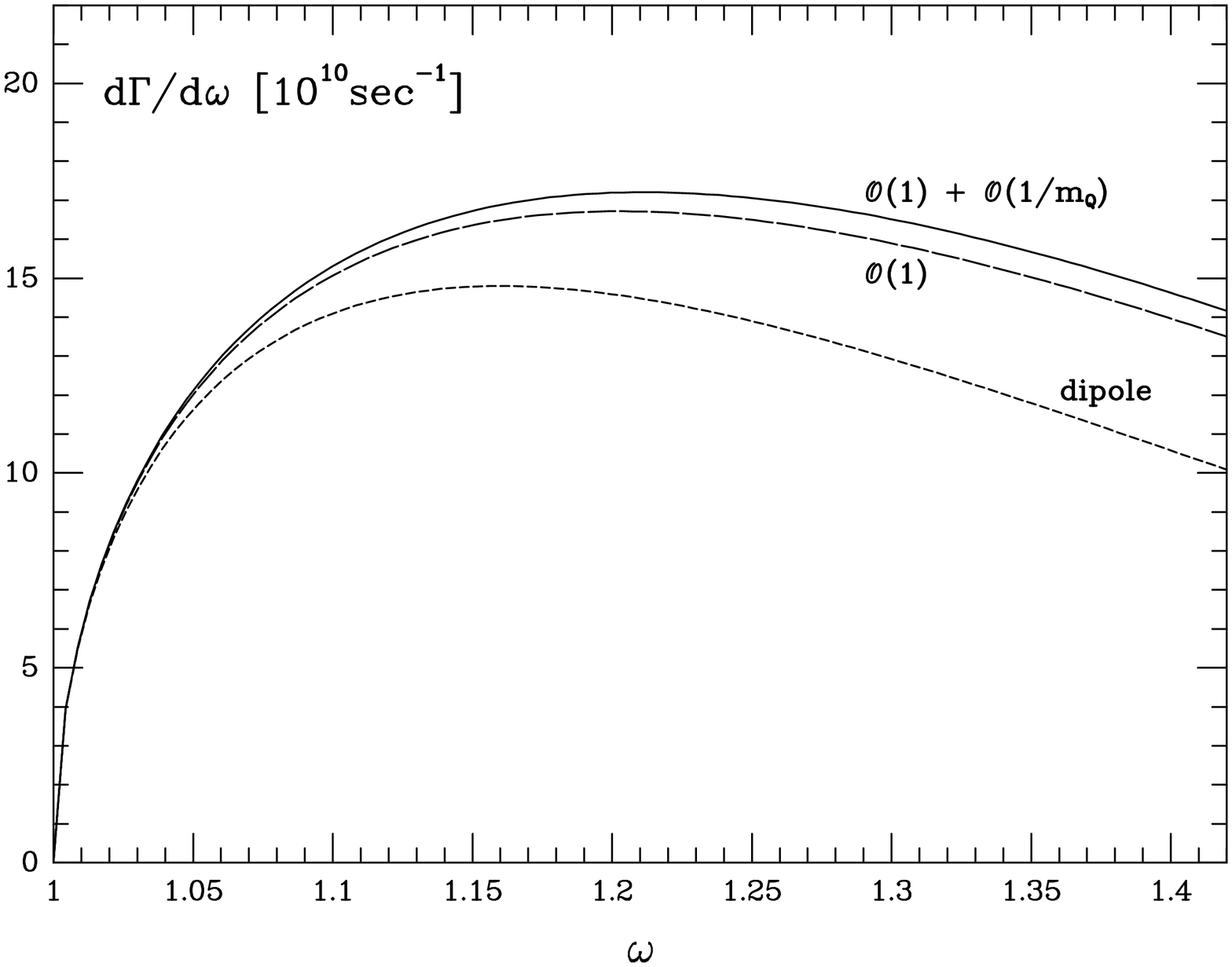,bbllx=10pt,bblly=60pt,bburx=550pt,bbury=770pt,%
        width=9.5cm,height=13cm,angle=-90}

\caption[dummy]{\label{fig_4} \it $\omega $ spectrum of decay rate in 
  the dipole model, the ${\cal O}(1)+{\cal O}(1/m_q)$ IMF quark model,
  and in the ${\cal O}(1)$ IMF quark model.}

\hspace*{12mm}
\epsfig{file=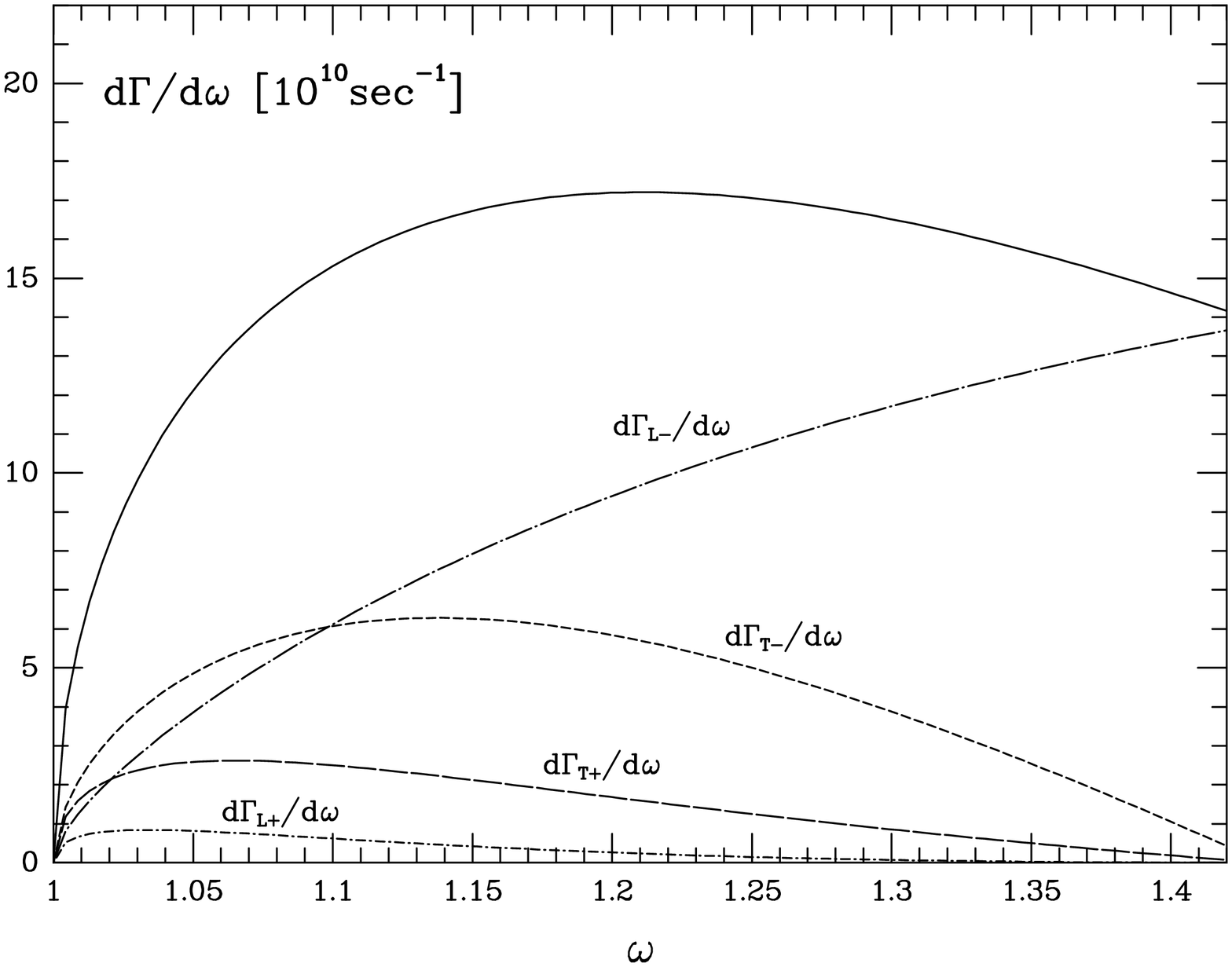,bbllx=10pt,bblly=60pt,bburx=550pt,bbury=770pt,%
        width=9.5cm,height=13cm,angle=-90}

\caption[dummy]{\label{fig_5} \it $\omega $ spectrum of decay rate and 
  partial rates into given helicity components for ${\cal O}(1) +
  {\cal O}(1/m_Q)$ IMF quark model.}

\end{figure}

In Fig.\ 5 we show the longitudinal/transverse composition of the
differential decay rate for the IMF quark model calculated up to
${\cal O}(1/m_Q)$.  The longitudinal rate $\Gamma_{L_-}$ (proportional
to $(\vert H_{-\frac{1}{2}0}\vert^2)$ dominates except at low
$\omega$. The left-chiral nature of the underlying $b \rightarrow c$
transition is reflected in the dominance of the transverse negative
rate $\Gamma_{T_-}$ over the transverse positive rate $\Gamma_{T_+}$
and the longitudinal negative rate $\Gamma_{L_-}$ over the
longitudinal positive rate $\Gamma_{L_+}$. This has interesting
experimental implications as will be discussed later on. As Fig.\ 6
shows the difference between the transverse and longitudinal negative
and positive rates is quite marked for high lepton momenta which are
best suited for experimental detection. 
\begin{figure}[t,h]

\hspace*{12mm}
\epsfig{file=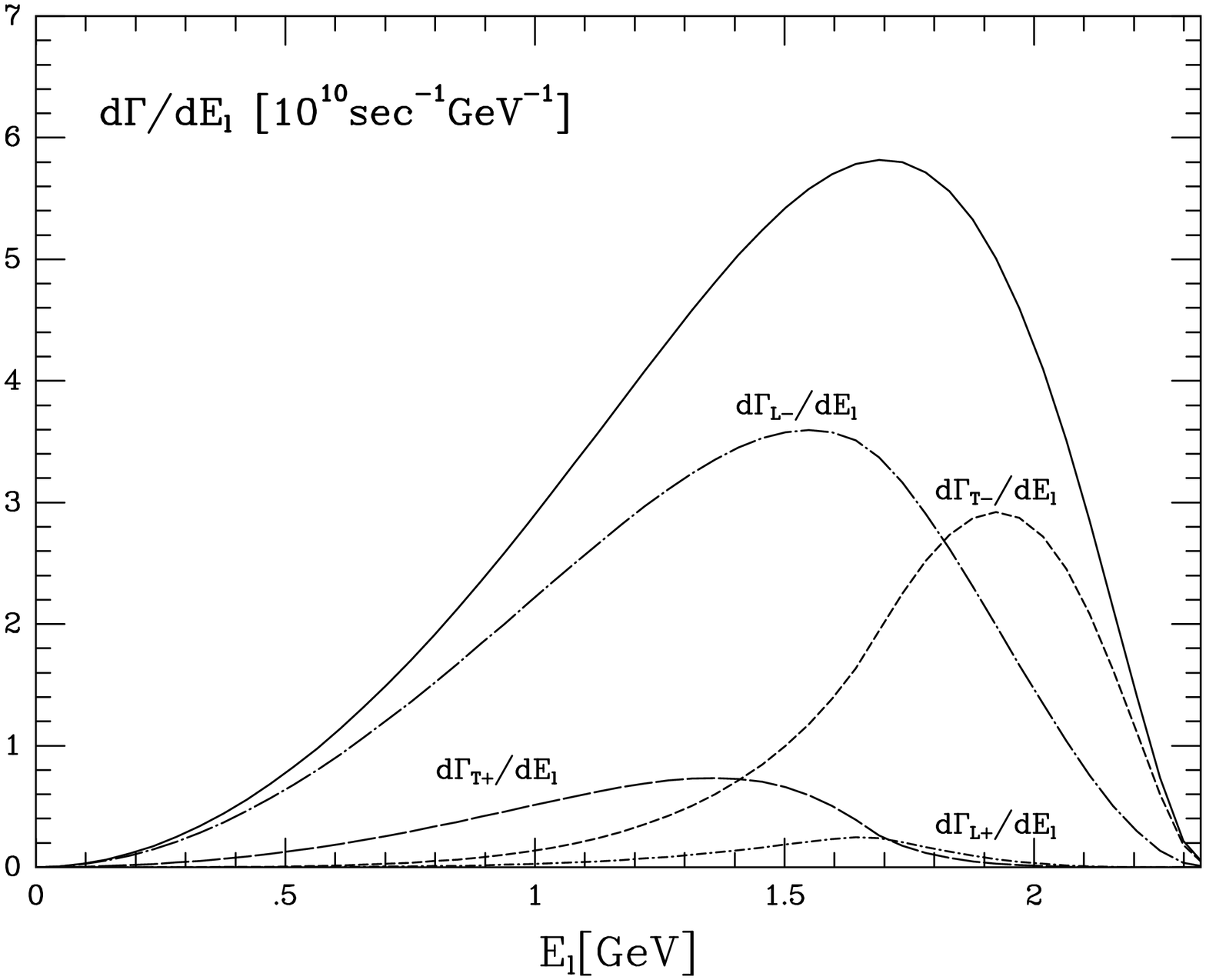,bbllx=10pt,bblly=60pt,bburx=550pt,bbury=770pt,%
        width=9.5cm,height=13cm,angle=-90}

\caption[dummy]{\label{fig_6} \it Lepton energy spectrum of decay rate 
  and partial rates into given helicity components for ${\cal O}(1) +
  {\cal O}(1/m_Q)$ IMF quark model.}

\end{figure}
 
The total decay rate $\Gamma_{tot}$ and the partial rates into given
helicity states of the $W^-$ (or the current) are listed in Table~1.
For the sake of comparison we also show results for the dipole model,
the free quark decay model and from an alternative IMF model
\cite{krollguo} which has many features in common with the model
presented in the present paper. We will comment on the model proposed
in \cite{krollguo} below. In all cases discussed in this paper the
longitudinal rate $\Gamma_{L_-}$ dominates over the transverse rates
$\Gamma_{T_+}$ and $\Gamma_{T_-}$ while the longitudinal positive rate
$\Gamma_{L_+}$ is small.  As expected from the left-handed current
coupling the transverse negative rate $\Gamma_{T_-}$ dominates over
the transverse positive rate $\Gamma_{T_+}$ and the longitudinal
negative rate $\Gamma_{L_-}$ dominates over the longitudinal positive
rate $\Gamma_{L_+}$.

\begin{center}
\begin{tabular}{|l|ccccc|}\hline
 & $\Gamma_{tot}$
 & $\Gamma_{T+}$
 & $\Gamma_{T-}$
 & $\Gamma_{L+}$
 & $\Gamma_{L-}$\\
\hline
${\cal O}(1)$ & 6.32 &  0.61  &   1.78 & 0.13 & 3.79 \\
${\cal O}(1) + {\cal O}(\frac{1}{m_Q})$
     & 6.50 &  0.62  &   1.82 & 0.14 & 3.92 \\
DIPOLE & 5.43 & 0.55   &   1.58 & 0.12 & 3.17 \\
FQD & 11.4 & 0.92   &   2.90 &  0.18 & 7.43 \\
\cite{krollguo} & 4.89  &   0.44  &   1.53  &  0.10 &  2.82 \\
\hline
\end{tabular}
\end{center}

\vspace*{1mm}

\noindent
{Table~1:} {\it
Total rates and partial rates into given helicity states.
Row 1: ${\cal O}(1)$ IMF quark model;
row 2: ${\cal O}(1)+{\cal O}(1/m_Q)$ IMF quark model;
row 3: dipole model;
row 4: free quark decay model (FQD) (or flat form factor model)
with $m_b = M_{\Lambda _b} = 5.621\,\,\mbox{{\rm GeV}}$ and
     $m_c = M_{\Lambda _c} = 2.285\,\,\mbox{{\rm GeV}}$;
row 5: ${\cal O}(1)+{\cal O}(1/m_Q)$ IMF quark model of {\rm \cite{krollguo}}.
Parameters for the IMF quark model are $m=n=1/2$,
$\alpha=0.5, \bar{\alpha}=0.2$~{\rm GeV}$^2$, $b=1.0$~{\rm GeV}$^{-1}$
and $\bar{b}=0.1$ Rates are given in units of $10^{10}
\,\mbox{{\rm sec}}^{-1}$.}


As already apparent from the differential rates Fig.\ 4 the IMF quark
model has the largest total rate $\Gamma_{tot}$; the ${\cal O}(1/m_Q)$
corrections are small and increase the ${\cal O}(1)$ rate by 3\%. It
is quite instructive to compare the computed rates with the free quark
decay (FQD) rates. For $m_b = 4.73\,\mbox{GeV}$ and $m_c =
1.55\,\mbox{GeV}$ one obtains $\Gamma_{tot} = 7.52 \times
10^{10}\mbox{s}^{-1}$. If one takes $m_b = 5.621 \, \mbox{GeV}$ and
$m_c = 2.285\,\mbox{GeV}$ (a choice which incorporates phases space
effects correctly) one finds $\Gamma_{tot} = 11.42\times
10^{10}\mbox{sec}^{-1}$. The latter case corresponds to taking
structureless form factors in the dipole model or in the ${\cal O}(1)$
HQET calculation.\footnote{Judging from the numbers in Table 1 the
  exclusive semileptonic decay rate $\Lambda_b \rightarrow
  \Lambda_{c}^{+} + l^- + \bar\nu_l$ would be predicted to amount to
  57\% - 87\% (IMF quark model) and 48\% - 72\% (dipole model) of the
  total inclusive semileptonic rate $\Lambda_b \rightarrow
  \Lambda_{c}^{+} + X + l^- + \bar\nu_l$.} The difference in rate
between the form factor models and the "structureless" rate
$\Gamma_{tot} \approx 7.52-11.42 \times 10^{10}\mbox{s}^{-1}$ would
have to be filled out by the contribution of higher $\Lambda_c$
resonances and continuum states.

The decay products in the quasi-two-body decay $\Lambda_b \rightarrow
\Lambda_{c}^{+} + W^-$ are highly polarized. The polarization of the
decay products can be analyzed by monitoring the angular decay
distributions of their subsequent decays. The structure of the
lepton-side decay $W^- \rightarrow l^- + \bar\nu_l$ is determined by
the Standard Model $(V - A)$ coupling and has 100\% analyzing power.
For the hadron side the two-body decay $\Lambda_{c}^{+} \rightarrow
\Lambda + \pi^{+}$ is best suited for this analysis since the decay
structure has recently been determined in two experiments \cite{27,28}
which obtained
\begin{equation}\label{alp}
\alpha_{\Lambda_c}
 = \bigg\{ \!\begin{array}{lc} -1.0 {+ 0.4 \atop - 0.0}& \cite{27}
        \\ -0.96 \pm 0.42 & \cite{28} \end{array}
\end{equation}
for the asymmetry parameter that characterizes the decay
$\Lambda_{c}^{+} \rightarrow \Lambda + \pi^+$.

The respective polar angle distributions are given by the following
expressions \cite{kk}
\begin{eqnarray}\label{polang}
\mbox{lepton side}&:& W(\Theta) = 1+2\alpha'\cos\Theta
+ \alpha''\cos^2\Theta \nonumber \\
\mbox{hadron side}&:& W(\Theta_\Lambda) = 1 + \alpha\alpha_{\Lambda_c}
\cos\Theta_\Lambda
\end{eqnarray}
where $\Theta$ and $\Theta_\Lambda$ are the polar angles of the lepton and
the $\Lambda$ in the $(l^-\overline{\nu}_l)$ CM system and the $\Lambda_c$
rest system, respectively (see \cite{kk}).
The asymmetry parameters in (\ref{polang}) can be expressed in terms of the
helicity amplitudes and read
\begin{equation}\label{12}
\alpha'=\frac{\hpl-\hmn}{\hpl+\hmn+2(\ho+\hoi)}
\end{equation}
\begin{equation}\label{13}
\alpha''=\frac{\hpl+\hmn-2(\ho+\hoi)}{\hpl+\hmn+2(\ho+\hoi)}
\end{equation}
\begin{equation}\label{10}
\alpha=\frac{\hpl-\hmn+\hoi-\ho}{\hpl+\hmn+\hoi+\ho}\quad .
\end{equation}
The asymmetry parameters $\alpha'$ and $\alpha''$ are specific
components of the polarization density matrix of the off-shell $W$,
whereas the asymmetry parameter $\alpha$ is the longitudinal
polarization $P_z$ of the daughter baryon $\Lambda_c$.  Mean values of
the above three asymmetry parameters are listed in Table~2. In
calculating the mean asymmetries one has to integrate numerator and
denominator separately. We also show results for asymmetry parameters
in Table~2 obtained from the dipole model, the free quark decay model
and from the alternative IMF model \cite{krollguo}.

\begin{center}
\begin{tabular}{|l|cccccc|}\hline
 & $\alpha$
 & $\alpha'$
 & $\alpha''$
 & $\gamma$
 & $\alpha_P$
 & $\gamma_P$ \\
\hline
${\cal O}(1)$ & $-0.76$ & $-0.11$ & $-0.53$ & 0.55  & 0.39  & $-0.16$ \\
${\cal O}(1) + {\cal O}(\frac{1}{m_Q})$ & 
     $-0.77$ & $-0.11$ & $-0.54$ & 0.55  & 0.40  & $-0.16$ \\
DIPOLE & $-0.75$ & $-0.12$ & $-0.51$ & 0.57  & 0.37  & $-0.17$ \\
FQD & $-0.81$ & $-0.10$ & $-0.60$ &  0.50 & 0.46  & $-0.14$ \\
\cite{krollguo} & $-0.78$ & $-0.14$ & $-0.49$ &  0.53 &  0.33 & $-0.15$ \\
\hline
\end{tabular}
\end{center}

\noindent
{Table~2:} {\it Mean values of various asymmetry parameters.
 Model parameters as in Table~1.}

\vspace*{2mm}

Alternatively one can define forward-backward asymmetries by averaging
over events in the respective forward (F) and backward (B) hemispheres
of the two decays and then by taking the ratio $A_{FB} = (F - B)/(F +
B)$. One then has
\begin{eqnarray}
\mbox{lepton side}&:& A_{FB} = -\frac{\alpha'}{1 + \frac{1}{3}\alpha''}
\nonumber \\
\mbox{hadron side}&:& A_{FB} = \frac{1}{2}\alpha\alpha_{\Lambda_{c}}
\end{eqnarray}
where the forward hemispheres are defined w.r.t. the momentum
direction of the $W^-$ and $\Lambda_b$, i.e. $\pi/2 \leq \Theta < \pi$
and $ 0 \leq \Theta_\Lambda < \pi/2$, respectively. Judging from the
large value of the measured asymmetry parameter $\alpha_{\Lambda_c}$
in Eq.\ (\ref{alp}) and the numbers in Table~2, the FB asymmetry on
the hadron side can be expected to be comfortably large in comparison
to the small FB asymmetry on the lepton side.

The FB asymmetry measures are quite interesting when one wants to
determine the chirality of the $b \rightarrow c$ transition. In the
left-chiral case, as predicted by the Standard Model, the $c$-quark,
and thereby the $\Lambda_c$ baryon, is predominantly in the negative
helicity state. As the asymmetry parameter $\alpha_{\Lambda_c}$ is
also negative, and thereby the helicity of the $\Lambda$ predominantly
negative, the helicities want to align, and one has an altogether
positive FB asymmetry. A right-chiral $b \rightarrow c$ transition
would, on the contrary, yield a negative FB asymmetry. The size of the
predicted FB asymmetry is large enough to accommodate even large
errors in this measurement to exclude or confirm the SM prediction for
the chirality of the $b \rightarrow c$ transition.

The FB asymmetry on the lepton side is again predicted to be positive
if the $b \rightarrow c$ transition is left-chiral. Again this can be
understood from simple helicity arguments. There are, however, two
reasons that the lepton side FB asymmetry measure is not optimal.
First, it is predicted to be quite small ($A_{FB} = 0.167$ in our
model), and second, one uses the left handedness of the lepton current
as input to analyze the chirality of the $b \rightarrow c$ coupling.
If the weak interaction were mediated by a non SM right-handed gauge
boson $W_R$ with right-handed couplings at the lepton {\it and} hadron
side one would have the same lepton-side FB asymmetry even though the
$b \rightarrow c$ transition is right chiral.\footnote{A viable model
  involving a right handed $W_R$ that is consistent with all present
  data has been recently proposed in Ref.\cite{growak}.}

The hadron-side FB asymmetry measure involves $P$-odd spin-momentum
correlations and thus is a direct measure of the $b \rightarrow c$
chirality whereas the lepton-side FB asymmetry involves $P$-even
momentum-momentum correlations only and is therefore not optimally
suited for the determination of the $b \rightarrow c$ chirality,
unless, of course, one presumes the handedness of $W^- \rightarrow
l^-\bar\nu_l$ is known.

In Table~2 we also list the value of the azimuthal asymmetry parameter
$\gamma$ which describes the azimuthal correlation of the lepton-side
and the hadron-side decay planes. The corresponding azimuthal
distribution is given by
\begin{equation}\label{14}
\frac{\mbox{d}\Gamma}{\mbox{d}q^2\mbox{d}\chi} \propto 1 -
\frac{3\pi^2}{32\sqrt{2}}\,\gamma\alpha_\Lambda\cos\chi
\end{equation}
where
\begin{equation}\label{15}
\gamma=\frac{2\mbox{Re}(H_{-\frac{1}{2}0}H_{\frac{1}{2}1}^{*}+
H_{\frac{1}{2}0}H_{-\frac{1}{2}-1}^{*})}{\hpl+\hmn+\ho+\hoi}
\end{equation}
and where $\chi$ is the relative azimuth of the two decay planes
(see \cite{kk}).

The asymmetry parameter $\gamma$ can be seen to be the transverse
component $P_x$ (in the lepton plane) of the polarization vector of
the daughter baryon $\Lambda_c$. Since we have taken the decay
amplitudes to be relatively real there is no $P_y$ component (out of
the lepton plane) and correspondingly no azimuthal term proportional
to sin$\chi$ in the angular decay distribution (\ref{14}).  The
presence of a $P_y$ polarization component would signal the presence
of CP-violating effects and/or final state interaction effects which
we shall not discuss in this paper.

For completeness we list in Table~2 also the values for the asymmetry
parameters $\alpha_p$ and $\gamma_p$ relevant for polarized
$\Lambda_b$ decays. They are related to polar and azimuthal
correlations between the polarization vector of the $\Lambda_b$ and
the momentum of the $\Lambda_c$ and the decay products of the
$\Lambda_c$ as described in \cite{kk}. An analysis of these decay
correlations is of relevance for $\Lambda_b$'s originating from
Z-decays which are expected to be produced with a substantial amount of
polarization \cite{close}. We must mention, though, that a first
analysis of the polarization of $\Lambda_b$'s from the Z  did not confirm 
theoretical expectations of a large $\Lambda_b$ polarization \cite{bus}.

Table~2 shows that the asymmetry parameters are not very dependent on
whether the IMF quark model or the dipole model is used as input
despite the fact that there are rate differences between the two. When
taking the asymmetry ratios differences in the $\omega$-dependence of
the form factors tend to drop out and one remains with the underlying
spin dynamics which is approximately the same in both models. Even
when going to the extreme case of choosing flat form factors for
$f_{1}^{V}$ and $f_{1}^{A}$ the asymmetry values do not change much
(see Table~2). We have checked that the same statement holds true when
one chooses $n=m=1$ for the endpoint power behaviour in the IMF quark
model.

Finally we remark that the form factors and the numerical rate values
presented in this paper are derived from unrenormalized current
vertices.  The renormalization effects can easily be incorporated as
discussed in detail in \cite{neuren} and in \cite{krollguo}. The
renormalization effects are very small close to the zero recoil point
$\omega = 1$ and become largest at maximum recoil $q^2 = 0$.
Numerically they tend to increase the rates by approximately 10\% but
leave the asymmetry values practically unchanged.

\section{Summary and Conclusions}
We have used a infinite momentum frame quark model that was improved
by using results from HQET to calculate the $\Lambda_b \rightarrow
\Lambda_c$ transition form factors and to give detailed predictions
for rates, spectra and polarization dependent observables in the
semileptonic decay $\Lambda_b \rightarrow \Lambda_{c}^{+} + l^- +
\overline{\nu}_l$ ($l = e, \mu$). We have employed heavy quark --
light diquark IMF wave functions in which the $x$-dependence of the
wave function resembles that of a mesonic heavy quark -- light
antiquark system. In our analysis we have only made use of the
so-called good components of the quark transition currents. 

It is important to realize that any BSW-type infinite momentum frame
quark model calculation does not take into account possible spin-spin
interactions between the heavy side and the light side as they occur
in general in a ${\cal O}(1/m_Q)$ HQET treatment. It is for this reason
that a calculation such as the one presented in \cite{nr} will lead to
inconsistencies when comparing the infinite momentum quark model
results with the general ${\cal O}(1/m_Q)$ HQET structure. In the case
of the calculation of \cite{nr} this inconsistency can be exhibited by
taking also the $B^* \rightarrow D^*$ channel into account, in
addition to the $B \rightarrow D,D^*$ treated in \cite{nr}. As
mentioned before there are no spin-spin interactions between the heavy
side and the light side in the case of the $\Lambda_b \rightarrow
\Lambda_c$ transitions and thus the infinite momentum frame structure
is fully consistent with the ${\cal O}(1/m_Q)$ HQET structure in this
special case.

At this point it is appropiate to summarize the uncertainties in our model 
predictions. The most important parameters are $n$ and $m$ controlling 
the powers of $x_1$ and $1-x_1$ in the wave function. Of less importance 
are the oscillator parameter $b$ and the mass difference $\alpha$ which 
determines the position of the maximum of the wave function. The 
${\cal O}(1)$ reduced form factor $F(\omega)$ depends only on $m$ and on 
the product of $\alpha$ and $b$ while the ${\cal O}(1/m_Q)$ form factor 
$\eta(\omega)$ depends on all parameters. 
While our analytical results are given for all values of $n$ and $m$
we have discussed two specific
choices of the parameters $n$ and $m$ in our numerical analysis , namely
$n=m=1/2$ and $n=m=1$. The choice $n=m=1/2$ is 
characterized by a relatively small value of the charge radius $\rho^2$ 
and, hence, a large exclusive semileptonic decay rate of the $\Lambda_b$
which will amount to a substantial fraction of the total inclusive 
semileptonic $\Lambda_b$ decay rate. Larger values of $m$ lead 
to larger values of the charge radius and thus to smaller exclusive 
semileptonic decay rates. Although we favor the choice $m=1/2$ 
the question of which value one finally has to choose for $m$
has to settled by experiment. There is first experimental
evidence that $n=m=1/2$ is the preferred choice \cite{Bertini}.
For a given value of $m$ the magnitude of the ${\cal O}(1/m_Q)$ reduced form 
factor and thereby the rate increases with $n$. Whereas the rates 
are strongly dependent on the choice of $n$ and $m$ our polarization
predictions show only a weak dependence on the 
choice of $n$ and $m$.

Before concluding this paper we would like to comment on an
alternative IMF quark model calculation of the $\Lambda_b \rightarrow
\Lambda_c$ transition form factors \cite{krollguo}. The IMF approach
of \cite{krollguo} has many features in common with the present model
calculation: The large form factors $f_{1}^{V}$ and $f_{1}^{A}$ are
calculated from overlap integrals of the initial and final state
hadronic wave functions for which the same quark-diquark light-cone
wave function is used as here. The other form factors $f_{2}^{V}$,
$f_{3}^{V}$, $f_{2}^{A}$ and $f_{3}^{A}$ are estimated in the same way
as we do here, namely from the HQET relations.

The difference between the two approaches is that in \cite{krollguo}
the overlap integrals are evaluated for all values of the momentum
transfer $q^2$ (or $\omega$) for given masses $M_c$, $M_b$ while in
this paper we use overlap integrals only at $q^2 = 0$ varying $\omega$
through the mass ratio $M_c/M_b$. As was mentioned above the overlap
integrals at $q^2 = 0$ are on rather save theoretical grounds, whereas
those at $q^2 \not= 0$ are somewhat less reliable. On the other hand,
the IMF that is being used here can not be reached from any frame of
finite momentum by a Lorentz transformation.

The results of the two models are, on the other hand, rather similar
mathematically as well as numerically (c.f.\ Tables~1,2). This gives us
additional confidence in our predictions. The main difference between
the predictions is that the form factors used in this paper exhibit a
singularity at $\omega = 0$ (see Eq.\ (\ref{itildedef})) which has no
physical interpretation while the form factor singularity in
\cite{krollguo} appears at $\omega = -1$ at physical threshold.
However, due to the normalization condition at zero recoil this
difference does not matter much numerically in the decay region.

\appendix
\section{Form factor relations including the bad current components}
In this Appendix we list the two additional equations relating form
factors and quark model overlap integrals when also the bad quark
current components are used. Using similar techniques as in the main
part of the paper one finds
\begin{eqnarray}\label{ffresult}
f_{1}^{V} (\omega, M_1, M_2) =
\frac{m_{1}-m_{2} }{ M_{1}-M_{2} }J(\omega,M_{1},M_{2})
\\ \nonumber
f_{1}^{A} (\omega, M_1, M_2) =
\frac{m_{1}+m_{2} }{ M_{1}+M_{2} }J(\omega,M_{1},M_{2})
\end{eqnarray}
where one now has a different overlap integral
\begin{equation}\label{Jover}
J(\omega, M_{1} , M_{2}) = \int_{0}^{1}{\rm d}x_{1} \phi_2^{*} (x_1)
                                \frac{1}{x_1}\phi_1 (x_1).
\end{equation}
It is clear that only the form factors $f_{1}^{V}$ and $f_{1}^{A}$
associated with the covariants $\gamma_\mu$ and $\gamma_\mu\gamma_5$
enter into the bad current relations since the covariants $v_{1\mu}$,
$v_{2\mu}$, $v_{1\mu}\gamma_5$ and $v_{2\mu}\gamma_5$ have no
transverse components.

Proceeding as in the main text we expand the overlap integral
$J(\omega, M_1 , M_2)$ according to
\begin{equation}\label{J}
J(\omega,M_1 ,M_2 ) = I^{(0)}(\omega)
                       \left( 1 + \left[ \frac{1}{M_1}+\frac{1}{M_2}
                         \right] \tilde{J}^{(1)}(\omega) + ...
                                                              \right)
\end{equation}
where
\begin{equation}\label{Jtildedef}
\tilde{J}^{(1)}(\omega) = \tilde{I}^{(1)}(\omega)
               + \frac{1}{2b} \frac{1}{\sqrt{\omega(\omega+1)}}
                 \frac{H_{2m+1}(\kappa\sqrt{\frac{\omega+1}{2\omega}})}
                 {H_{2m}(\kappa\sqrt{\frac{\omega+1}{2\omega}})}.
\end{equation}
Note that the zeroth order coefficient of $J(\omega)$ is identical to
the zeroth order coefficient $I^{(0)}(\omega)$ (see Eq.\ (\ref{Itilde})).
Similarly we expand the mass factors in Eq.\ (\ref{ffresult}) up to
${\cal O}(1/M_Q)$ using $m_i + \alpha_i = M_i$. One obtains
\begin{equation}\label{weissnich}
\frac{m_1 + m_2}{M_1 + M_2} = 1 - \left ( \frac{1}{M_1} + \frac{1}{M_2}
\right ) \frac{\alpha}{\omega + 1}; \quad \quad \quad
\frac{m_1 - m_2}{M_1 - M_2} = 1.
\end{equation}

In Sec.\ 3 we have argued that the relations (\ref{ffresult}) obtained
from the bad current components are not so reliable. One can
nevertheless ponder the question which kind of relations one obtains
for the form factors if the combined set of four equations
(\ref{0and3a},\ref{0and3b}) and (\ref{ffresult}) is used as a starting
point. In particular one can ask oneself whether one can now derive
the HQET relations Eqs.\ (\ref{fivfiaone},\ref{kkp}) or
(\ref{georgi}).

At ${\cal O}(1)$ one derives from the bad current relations Eq.\ 
(\ref{ffresult}) $f_{1}^{V\, (0)}(\omega) = f_{1}^{A\, (0)}(\omega) =
I^{(0)}(\omega) = F(\omega)$ where the zero recoil condition
$f_{1}^{A\, (0)}(\omega = 1) = 1 $ is satisfied.  Substituting this
result into the good current relations Eq.\ 
(\ref{0and3a},\ref{0and3b}) one finds $f_{2}^{V\, (0)}(\omega) M_2 +
f_{3}^{V\, (0)}(\omega) M_1 = 0$ and $f_{2}^{A\, (0)}(\omega) M_2 +
f_{3}^{A\, (0)}(\omega) M_1 = 0$.  Further one has the charge
conjugation relations $f_{2}^{V\, (0)}(\omega) = f_{3}^{V\,
  (0)}(\omega)$ and $f_{2}^{A\, (0)}(\omega) = -f_{3}^{A\,
  (0)}(\omega)$ which then imply that $f_{2}^{V\, (0)} = f_{3}^{V\,
  (0)} = f_{2}^{A\, (0)} = f_{3}^{A\, (0)} = 0 $. This then
establishes that the ${\cal O}(1)$ HQET result Eq.\ (\ref{fivfiaone})
can in fact be derived when both the good and bad current quark model
relations are used.

Turning now to the ${\cal O}(1/m_Q)$ results, the bad current relation
Eq.\ (\ref{ffresult}) immediately leads to the first of the HQET
results in Eq.\ (\ref{georgi}) upon using the expansion
(\ref{weissnich}). As for the good current relations, one has to 
identify the HQET parameter $\bar{\Lambda}$ with the quark model 
parameter $\alpha$, i.e.\ $\bar{\Lambda} = \alpha$.

The reduced form factor $\eta(\omega)$ (see (\ref{kkp})) can also be
calculated from the bad current relations (\ref{ffresult}) upon using
the expansion (\ref{weissnich}) again. In fact one finds
\begin{equation}
  \label{etab}
  \eta_{b.c.}(\omega) = I^{(0)}(\omega) \left[
                       2 \tilde{J}^{(1)}(\omega) -
                       \bar{\Lambda} \right]
\end{equation}
which differs from the solution (\ref{solution}) obtained from the
good current relations. 
While the reduced form factor $\eta_{b.c.}$ is fairly small 
in the $\Lambda_b \rightarrow \Lambda_c$ decay region  as compared to
$F(\omega)$ it is not zero at $\omega = 1$.
With regard to the approximations involved in the bad current
components (see the discussion in Sect.\ 3) this little inconsistency 
in the ${\cal O}(1/m_Q)$ results can be tolerated allowing us to 
conclude that even the bad current relations are in reasonable 
agreement with HQET.

\end{document}